\documentclass[epsf]{aastex}
\usepackage{apjfonts}
\usepackage[onecolumn]{emulateapj5}

\newcommand{\sfr}{\,{\rm M_\odot\,yr^{-1}}}
\newcommand{\kms}{\,{\rm km\,s^{-1}}}
\newcommand{\bea}{\begin{eqnarray}}
\newcommand{\ena}{\end{eqnarray}}
\newcommand{\bec}{\begin{center}}
\newcommand{\enc}{\end{center}}

\begin{document}

\title{The State of the Circumstellar Medium Surrounding Gamma-Ray
Burst Sources and its Effect on the Afterglow Appearance}

\author{Enrico Ramirez-Ruiz\altaffilmark{1,}\altaffilmark{2},
Guillermo Garc\'{\i}a-Segura\altaffilmark{3} , Jay
D. Salmonson\altaffilmark{4} , and Brenda
P\'erez-Rend\'on\altaffilmark{3,5}}

\altaffiltext{1}{Institute for Advanced Study, Einstein Drive,
Princeton, NJ 08540: enrico@ias.edu} \altaffiltext{2}{Chandra Fellow}
\altaffiltext{3}{Instituto de Astronom\'ia-UNAM, Apartado Postal 877,
Ensenada, 22800 Baja California, Mexico: ggs@astrosen.unam.mx.}
\altaffiltext{4}{Lawrence Livermore National Laboratory, Livermore, CA
94551: salmonson@llnl.gov }
\altaffiltext{5}{Universidad de Sonora, Hermosillo, Sonora, Mexico}

\begin{abstract}
We present a numerical investigation of the contribution of the
presupernova ejecta of Wolf-Rayet stars to the environment surrounding
gamma-ray bursts (GRBs), and describe how this external matter can
affect the observable afterglow characteristics. An implicit
hydrodynamic calculation for massive stellar evolution is used here to
provide the inner boundary conditions for an explicit hydrodynamical
code to model the circumstellar gas dynamics.  The resulting
properties of the circumstellar medium are then used to calculate the
deceleration of a relativistic, gas-dynamic jet and the corresponding
afterglow light curve produced as the shock wave propagates through
the shocked-wind medium. We find that variations in the stellar wind
drive instabilities that may produce radial filaments in the
shocked-wind region. These comet-like tails of clumps could give rise
to strong temporal variations in the early afterglow lightcurve.
Afterglows may be expected to differ widely among themselves,
depending on the angular anisotropy of the jet and the properties of
the stellar progenitor; a wide diversity of behaviors may be the rule,
rather than the exception.
\end{abstract}


\section{Introduction}
\label{int}
Over the past six years evidence has mounted that long-duration ($\ge
2$ s) gamma-ray bursts (GRBs) signal the collapse of massive stars in
our Universe. This evidence was originally based on the probable
association of the unusual GRB 980425 with a type Ib/c supernova (SN;
Galama et al. 1998) but now includes the association of GRBs with
regions of massive star formation in distant galaxies (Paczy\'nski
1998; Wijers et al. 1998; Fruchter et al.  1999; Djorgovski
et~al. 2001; Trentham et~al. 2002), the appearance of supernova-like
{\it bumps} in the optical afterglow light curves of several bursts
(Bloom et al. 1999; Zeh, Klose \& Hartmann 2004 and references
therein), lines of freshly synthesized elements in the spectra of a
few X-ray afterglows (Piro et al. 2000; Ballantyne \& Ramirez-Ruiz
2001; Reeves et~al. 2002), and the first convincing spectroscopic
evidence that a very energetic supernova (a hypernova) was temporally
and spatially coincident with a GRB (Hjorth et al. 2003; Stanek et
al. 2003). These observations support the idea that long-duration GRBs
are associated with the deaths of massive stars, presumably arising
from core collapse (Woosley 1993, Zhang et al. 2003).

An implication of a massive star progenitor is that the circumburst
environment is determined by the mass-loss wind from the star.  Much
of our effort in this paper will therefore be dedicated to determining
the state of the circumburst material in various types of progenitor
scenarios, and describing how this external matter can affect GRB jets
propagating through it.  It should be noted here that the afterglows
sample a region $\sim 10^{17}$ cm in size. Because massive stars are
expected to have their close-in surroundings modified by the
progenitor winds, we consider both free winds and shocked winds as
possible surrounding media for the afterglow stage.  Detailed
hydrodynamic simulations of this interaction are presented in \S
4. Our computations allow us, for the first time, to study the
behaviour of circumstellar gas very close to the progenitor star. An
understanding of the evolution of a gas-dynamic jet can come only
through a knowledge of the properties of the medium which it
propagates. Calculations of the evolution of a relativistic,
gas-dynamic jet and expected emission properties are discussed in \S
5. For completeness, the interactions with either a free wind or a
constant density medium, as well as the termination shock wave which
marks the transition between these two media, are discussed in \S 2
and \S 3. The role of binarity is briefly addressed in \S 4.1. We then
in \S 6 discuss the possible variety of afterglow variability that is
expected from GRB jets expanding in a medium that may be inhomogeneous
because of clumping in the Wolf-Rayet (WR) star wind. Discussion and
conclusions are presented in \S 7.

\section{The Circumburst Medium}
If the progenitors are massive stars then there is an analogy to the
explosions of core collapse supernovae, for which there is abundant
evidence that they interact with the winds from the progenitor
stars. In most supernova cases, the radial range that is observed is
only out to a few pc, such that the mass loss characteristics have not
changed significantly during the time that mass is supplied to the
wind (Chevalier \& Li 2000). The density in the wind depends on the
type of progenitor. Red supergiant stars, which are thought to be the
progenitors of Type II supernovae, have slow dense winds. Wolf-Rayet
stars, which are believed to be the progenitors of Type Ib/c
supernovae and possibly of GRBs (e.g. MacFadyen \& Woosley 1999), have
faster, lower-density winds. The winds from typical red supergiants
are slow-moving and dense, with velocities $v_w\approx 10-20$ km
s$^{-1}$ and mass loss rates between $10^{-6}$ and $10^{-4}\,\sfr$
(e.g. Fransson et al. 1996). The winds from WRs, on the other hand,
are characterized by mass-loss rates $\dot M\approx 10^{-5}-10^{-4}
\,\sfr$ and velocities $v_w\approx 1000-2500\,\kms$ (e.g. Chiosi \&
Maeder 1986). In a steady, spherically symmetric wind, the electron
density is
\begin{equation}
n_w(r)={\dot M \over 4\pi v_w r^2 \mu_e m_p} \approx 3\times
10^6\,{\rm cm}^{-3} r_{15}^{-2} \dot M_{-4} v_{w,3}^{-1} \mu_e^{-1},
\label{eqn:wrwinds}
\end{equation} 
where $\mu_e$ is the molecular weight per electron and $\mu_e\sim 2$
in a helium gas. Here $v_w=10^3 v_{w,3} \kms$, $r=10^{15}r_{15}$ cm,
and $\dot M=10^{-4}M_{-4}\sfr$.

For this discussion we shall assume the blast wave is adiabatic,
i.e. its energy is constant with time, and effectively spherical. This
means the $E$ here is the isotropic equivalent energy as, for example,
derived from the gamma-ray output.  Deceleration due to the stellar
wind starts in earnest when about half the initial energy is
transferred to the shocked matter, i.e. when it has swept up
$\gamma^{-1}$ times its own rest mass. The typical mass where this
happens is
\begin{equation}
M_{\rm dec}={E \over \gamma^{2}c^2} \approx 5 \times 10^{-6}
E_{53}\gamma_{2}^{-2} M_\odot.
\label{eqn:mdec}
\end{equation}
The relativistic expansion is then gradually slowed down, and the
blast wave evolves in a self-similar manner with a power-law
lightcurve. This phase ends when so much mass shares the energy that
the Lorentz factor, $\gamma$, drops to 1. Obviously, this happens when
a mass $E/c^2$ has been swept up. This sets a non-relativistic mass
scale
\begin{equation}
M_{\rm NR}={E \over c^2} \approx 6 \times 10^{-2} E_{53} M_\odot.
\label{eqn:mnr}
\end{equation} 
Beyond this point, the event slowly changes into a classical
Sedov-Taylor supernova remnant evolution, leading to a steeper decline
in the lightcurve (Waxman et al. 1998).

\begin{figure}[a]
\centering \noindent {\par\centering
\resizebox*{0.7\columnwidth}{!}{\includegraphics{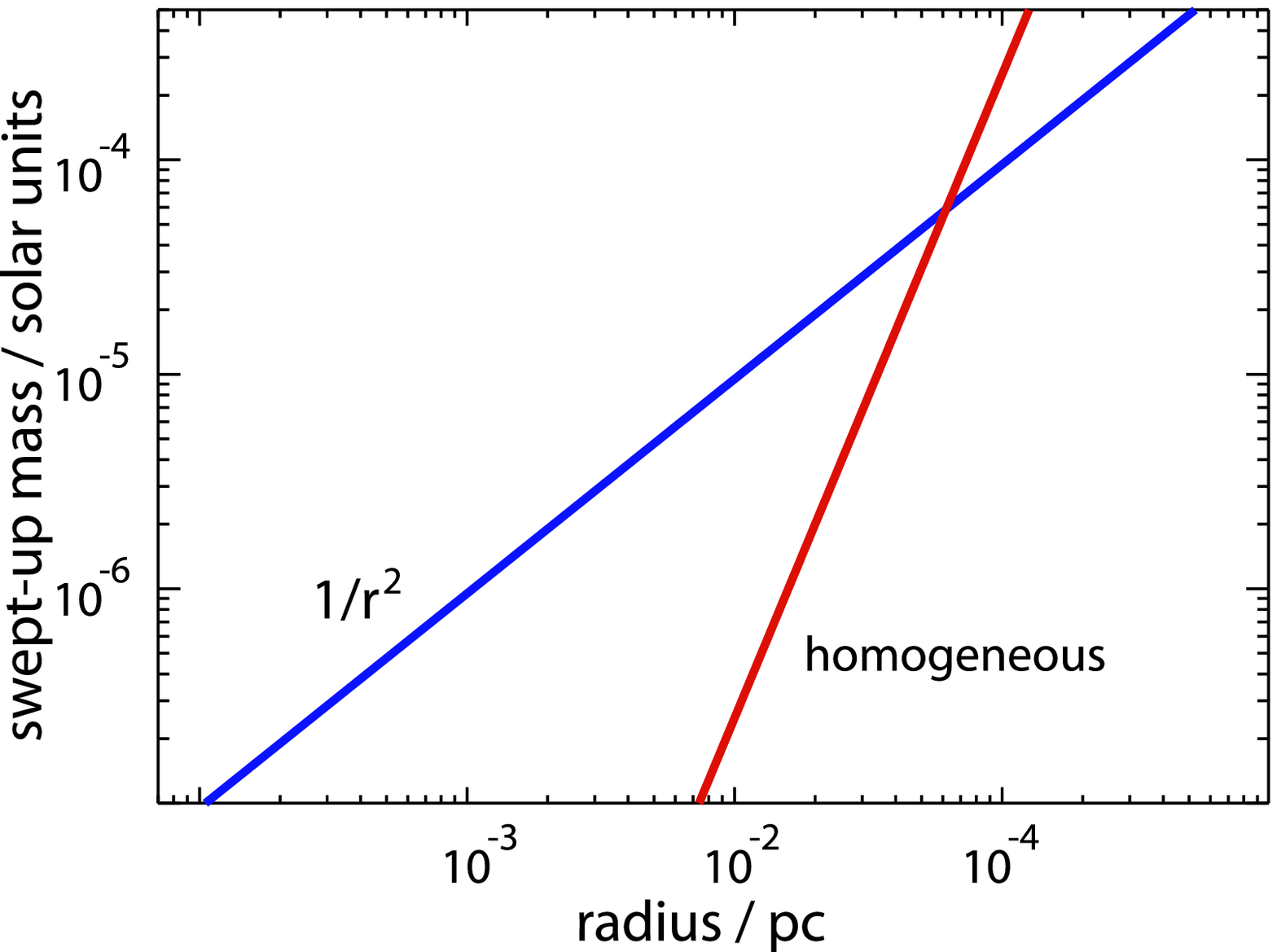}} \par}
\caption{Constraints on the swept-up mass $M$ as a function of
radius. The blue line is the swept-up wind mass assuming
$\dot{M}=10^{-6} M_\odot {\rm yr}^{-1}$ and $v_w\sim 10^{3} {\rm
km\;s}^{-1}$. The red line, on the other hand, assumes a uniform
medium with 1 cm$^{-3}$.}
\label{fig:am}
\end{figure}

In the unshocked wind, the mass within radius $r$ is $\dot{M}r/v_w$,
which combined with equation (\ref{eqn:mdec}) gives the blast wave
deceleration radius in a stellar wind:
\begin{equation}
r_{\rm d}^{w}={Ev_w \over \dot{M}c^2\gamma^2} \approx 2 \times
10^{16}E_{53}v_{w,3} \dot{M}_{-6}^{-1} \gamma_2^{-2} {\rm cm},
\label{eqn:rwdec}
\end{equation} 
where $E_{53}=(E/10^{53})$ ergs. By contrast, the well-known
expression for the uniform-medium deceleration radius is
\begin{equation}
r_{\rm d}\approx 10^{18}(E_{53}/n_{{\rm ism},0})^{1/3}\gamma_2^{-8/3}
{\rm cm},
\label{eqn:rdec}
\end{equation} 
and so a blast wave in a wind decelerates at a much smaller
radius. The deceleration time is given by $t_{\rm d}=r_{\rm
d}/(2\gamma^2c)$ in both cases, and thus is correspondingly smaller
for the wind case. However, because the Lorentz factor decreases as
$M^{-1/2}$ (with $M$ being the swept-up mass) beyond this point, the
mass-starved wind blast wave decelerates much more slowly, and
therefore begins to catch up. There is therefore not much difference
in size between wind and uniform CM blast waves in the most commonly
observed interval from 0.3 to 10 days after the burst
(Fig. \ref{fig:am}). Usually in the study of afterglows, one considers
either the uniform ambient medium case or the $1/r^2$ wind case on its
own\footnote{Although the interstellar and wind models are the two
main types of environments considered for afterglows, there is a
different scenario involving a massive star in which the supernova
explosion occurs before the GRB (Vietri \& Stella 1998). The supernova
would expand into the progenitor wind, creating a complex circumburst
region in the inner part of the wind. Konigl \& Granot (2002) have
recently shown, for the case of a pulsar-wind bubble, that the shocked
wind has a roughly uniform density, similar to that found in the
normal interstellar medium.}. However, since the wind of a star meets
the interstellar medium (ISM) at some point, the density structure is
more complex, and it is to this problem that we now turn our
attention.

\section{Wind-ISM Interaction}
\begin{figure}[a]
\centering \noindent {\par\centering
\resizebox*{0.85\columnwidth}{!}{\includegraphics{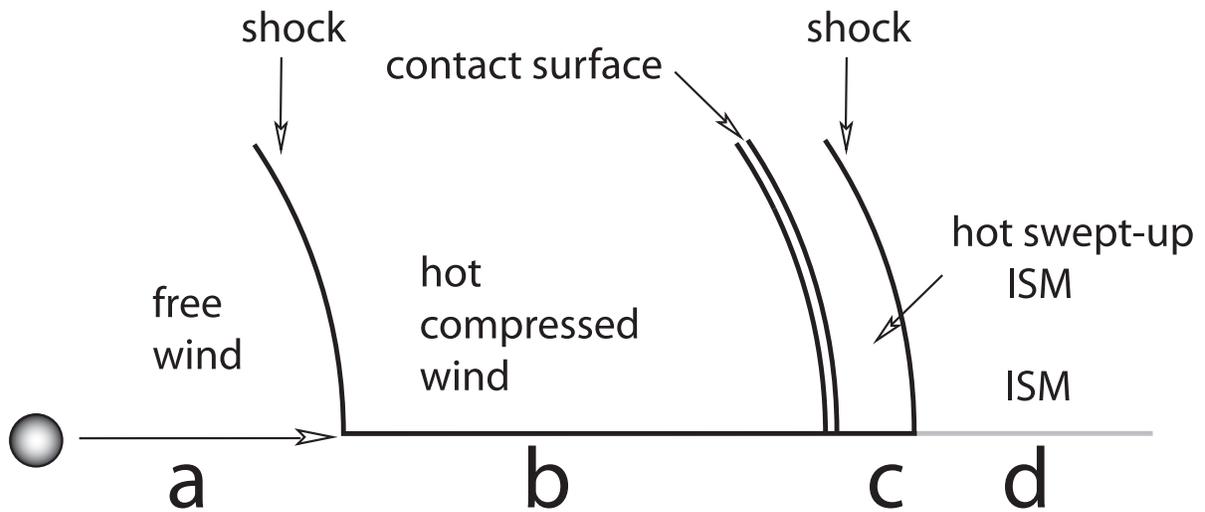}} \par}
\caption{The formation of shocks by a stellar driven wind interacting
with the surrounding interstellar medium. The interaction leads to a
driven wave composed of shock compressed wind, a contact surface, and
the swept-up ISM. Fast stellar wind matter (zone [a]) enters the shock
where it is compressed by a factor of $\sim 4$, and heated; the
accumulated shock wind material is in zone (b). The entire region
between the two shocks is nearly isobaric, so if material cools, it
becomes compressed and resides in the contact surface, which forms a
boundary between the shocked wind and the shocked, swept-up ISM (zone
[c]). The undisturbed ISM is in zone (d).}
\label{fig:sd}
\end{figure}

During the evolution of a wind-driven circumstellar shell the system
has a four-zone structure (analogous to that of a supernova shell;
Woltjer 1972). From the inside to the outside these zones are: (a) a
supersonic stellar wind with density $\rho (r)=\dot{M}/ 4\pi r^2
v_{w}$; (b) a hot, almost isobaric region consisting of shocked
stellar wind mixed with a small fraction of the swept-up interstellar
gas; (c) a thin, dense, cold shell containing most of the swept-up
interstellar gas; (d) ambient interstellar gas of number density
$n_{\rm ism}$ (Fig. \ref{fig:sd}).\\

The wind initially expands unopposed into the ISM with a velocity of
about $v_w \sim v_{\infty}$, the escape velocity at the sonic point.
The free expansion phase is considered to be terminated at a time
$t_{w,{\rm d}}$, when the swept-up mass of the interstellar medium is
comparable to the mass in the wind. The mass lost by the star is
$\dot{M}t_{w,{\rm d}}$ and the swept-up mass is $ {4\pi \over
3}(v_{w}\;t_{w,{\rm d}})^3 n_{\rm ism} m_{p} \mu_e$.  These two
masses are equal when $t_{w,{\rm d}}=[3\dot{M}/ (4 \pi v_{w}^3 n_{\rm
ism} m_{p} \mu_e) ]^{1/2}$, which is about 100 years for a typical WR
wind expanding into an homogeneous ISM. The free-expansion phase takes
place at the early stages of the evolution of the hot star and
occupies a minimal fraction of its lifetime. During this time both
$\dot{M}$ and $v_{w}$ are approximately constant, and the wind bubble
has reached a radius of $\sim 9 \times 10^{17} \dot{M}_{-5} n_{{\rm
ism},1}^{1/2}v_{w,3}^{-1/2} {\rm cm}$, where $\dot{M}$ is the mass
loss rate in units of solar masses per year, and $n_{\rm ism}$ is the
density of the surrounding medium in units of cm$^{-3}$.

When the free-expansion phase (a) has ended, the wind encounters an
inward facing shock.  Kinetic energy is deposited in the shocked wind
region in the form of heat,
   \begin{equation}
T_{\rm shock}= {3 \over 16}{m_{p} \mu_e \over k}(\Delta v_w)^2= 1.4
\times 10^5 ({\Delta v_w \over 100\;{\rm km\;s}^{-1}})^2 \;{\rm K},
\label{eqn:tshock}
\end{equation}
where $\Delta v_w$ is the relative speed of the material approaching
the shock. Thus, a jump in velocity by $800$ km s$^{-1}$, which is
still well below typical terminal wind speeds, produces a $10^{7}$ K
gas in the shocked wind region. During phase (b), the material is so
hot that it causes the contact surface to expand outward more slowly
than it would in a freely expanding wind.  The ISM that enters the
outward facing shock is heated to a temperature below $10^6 {\rm K}$,
emission of line radiation becomes the dominant cooling process and
the swept-up gas cools quickly to temperatures of about $10^4 {\rm K}$
that can be maintained by the radiation field of the star. The
duration of the adiabatic expansion phase can thus be estimated by
finding the time it takes the expanding gas to cool from $T_{\rm
shock} \approx 10^{7} {\rm K}$ to $10^{6} {\rm K}$. Using equation
(\ref{eqn:tshock}), we find that a change in temperature from $10^{7}
{\rm K}$ to $10^{6} {\rm K}$ corresponds to a change in jump velocity
by a factor of $\sqrt{10}$.  This change in jump velocity corresponds
to a phase (b):(a) age ratio of about~6 (Castor et al. 1975). Thus,
the age of the adiabatic phase is less than about 1000 years.

The mass of the swept-up material is much larger than that in the hot
wind and, because it is cool, it lies in a compressed region.  Phase
(c) persists for as long as the star is able to sustain a powerful
wind. The dominant energy loss of region (b) is work against the
compressed region (c). The compressed region (c) expands because its
gas pressure is higher than that of the surrounding ISM. Therefore,
the expansion is described by the momentum equation,
\begin{equation}
{d \over dt}[{M_{\rm S}(t)v(t)}]=4\pi r^2(t) P_{\rm i},
\label{eqn:momeq}
\end{equation}
where $P_{\rm i}$ is the internal pressure of the compressed region,
assuming that most of the swept-up interstellar mass remains in the
thin shell. $M_{\rm S}(t)$ is the mass of the shell of swept-up
material, given by $M_{\rm S}(t)=(4/3)\pi r^3(t) \rho_{\rm
i}$. $P_{\rm i}$ is determined by the gas pressure of the high
temperature gas in region (b). The wind material that enters the
backward-facing shock is hot, but the material that enters the
forward-facing shock is cool. The cooled swept-up material is driven
outward by the high gas pressure of the hot bubble. The stellar wind
adds energy to region (b) at a rate
\begin{equation}
L(t)={1 \over 2}\dot{M}(t) v_{w}^2(t).
\end{equation}
The internal energy in the bubble is given by the product of the
energy per unit mass of the material, $(3/2)nkT / \rho_{\rm
i}=(3/2)P_{\rm i}/ \rho_{\rm i}$, and the total mass of the bubble,
$(4/3) \pi r^3 \rho_{\rm i}$. Since the total internal energy of the
bubble comes from the energy of the wind, we find $\dot{P}_{\rm
i}=L(t) / [2 \pi r^3(t)]$.

The expansion of the bubble during the adiabatic phase can be found
numerically by using this expression in the momentum equation. If the
wind power $L$ is roughly constant for a period of time, $t$, one can
write $P_{\rm i}={L\;t / (2 \pi r^3)}$. The resulting solution of
equation (\ref{eqn:momeq}) gives $r(t) \propto t^{3/5}$. This shows
that the shell expands more slowly than would a freely expanding
wind. Since the gas in the cavity moves subsonically its pressure
keeps it approximately at uniform density. The bubble could continue
to expand until stalled by the pressure of the ISM (Garc\'{\i}a-Segura
\& Franco 1996).

The radius of the wind termination shock at the inner edge of the wind
bubble can be found by balancing the wind ram pressure with the
post-shock cavity pressure. For a star that loses mass at a rate
$10^{-6}\dot{M}_{-6} M_\odot {\rm yr}^{-1}$ with a wind velocity
$10^{3}v_{w,3}$ km s$^{-1}$ in interstellar gas with density $10^{3}
n_{{\rm ism},3} {\rm cm}^{-3}$, we have a inner termination shock
radius
\begin{equation}
r_t(t)=0.4 \dot{M}_{-6}^{3/10} v_{w,3}^{1/10} n_{{\rm ism},3}^{-3/10}
t_6^{2/5} \rm{pc},
\end{equation}
where $10^{6}t_6$ is the lifetime of the star in Myr. The density in
the uniform shocked wind region, $n_{\rm sw}$, at late times is given
by
\begin{equation}
n_{\rm sw} \sim {3 \dot{M} \over {4\pi r_t^2 v_{w} m_p}} = 0.06
\dot{M}_{-6}^{2/5} n_{{\rm ism},3}^{3/5} v_{w,3}^{-6/5} t_6^{-4/5}
{\rm cm^{-3}},
\end{equation}
which shows that even if the progenitor star is embedded in a dense
molecular cloud the observed blast wave can propagate in a
low-density, uniform medium (Wijers 2001). The mass within the
$1/r^{2}$ wind, $M_t$, is
\begin{equation}
M_t = { \dot{M} r_t \over v_{w}} = 3 \times 10^{-4} \dot{M}_6^{13/10}
v_{w,3}^{-9/10} n_{{\rm ism},3}^{-3/10} t_6^{2/5} M_{\odot}.
\end{equation}
Comparison with estimates given in Soderberg \& Ramirez-Ruiz (2002)
show that if the wind is particularly weak (i.e. $\dot{M} \leq
10^{-6}M_\odot {\rm yr}^{-1}$) or the surrounding density is high
($n_{\rm ism} \geq 10^{3}\;{\rm cm}^{-3}$), $r_t$ falls within the
range of the relativistic expansion. Models and observations of
Galactic Wolf-Rayet stars, however, show that the swept-up shell of a
red supergiant material at the outer radius is at a distance $\ge 3$
pc from the star (Garc\'{\i}a-Segura et al. 1996a). This radius is
sufficiently large that the interaction with the free $1/r^2$ wind is
expected over the typical period of observation of afterglows.

Among the afterglows that can be interpreted as interaction with a
undisturbed stellar wind, the highest density objects are compatible
with expectations for the wind from a typical Wolf-Rayet star
(Panaitescu \& Kumar 2002), but the lower densities imply a wind
densities that are lower by a factor of $10-10^2$. As proposed by
Wijers (2001), the low-density requirement may be alleviated by
appealing to a shocked wind, but the observable afterglow transitions
due to the blast wave traversing the wind termination shock wave
(Fig. \ref{fig:sd}) have not been seen in any afterglow.

Depending upon the wind history of a massive star during its last few
centuries, however, the density structure in this inner region
(i.e. relevant to the afterglow phase) could be quite complicated as
the star enters advanced burning stages unlike those in any Wolf-Rayet
star observed so far. The non-steady nature of the winds in massive
stars therefore leaves open the possibility of interaction with denser
material at early times (Chevalier et al. 2004). This encourages us to
present a detailed account of the large and small scale features that
may be present in the circumstellar gas distribution.  These features,
as we argue, can result naturally when one takes into account the
complete mass-loss history of a massive star.

\section{Wind-Wind Interaction}
The detailed dynamical evolution of the circumstellar material (CSM)
around massive stars is complex. Some stages of it do not involve
major hydrodynamical instabilities and can thus be studied
analytically by means of self-similar solutions (Weaver et al. 1977;
Ostriker \& McKee 1988; Chevalier \& Liang 1989 ; Garc\'{\i}a-Segura
\& Mac Low 1995a). However, the frequent occurrence of instabilities
requires two -- or three -- dimensional hydrodynamic calculations in
order to follow the non-linear evolution of the resulting structures
(Franco et al. 1991; Blondin \& Lundquist 1993; Garc\'{\i}a-Segura \&
Mac Low 1995b).

In an effort to achieve full consistency between stellar and
circumstellar evolution, we have performed several computations where
the time-dependent input for the calculation of the circumstellar gas
dynamics is derived from the output of a stellar evolution code
developed at the University of G\"ottingen (Langer et al. 1988, Langer
1991). Here we carry out computational simulations with the hydrocode
ZEUS-3D (version 3.4) developed by M. L. Norman and the Laboratory for
Computational Astrophysics. ZEUS-3D is a finite-difference, fully
explicit, Eulerian code (Clarke 1996) descended from the code
described by Stone \& Norman (1992).  We used spherical coordinates
for our simulations, with a symmetry axis at the pole, and reflecting
boundary conditions at the equator and the polar axis. The reader is
referred to Garc\'{\i}a-Segura et~al. (1996a,b) for a review of the
applied computational methods and techniques.

The preburst stellar wind depends on the evolutionary stages prior to
(and during) the Wolf-Rayet stage. For Galactic stars, a standard
evolutionary track is to start as an O star, evolve through a red
supergiant (RSG) phase or luminous blue variable (LBV) phase with
considerable mass loss, and ending as a Wolf-Rayet star
(Garc\'{\i}a-Segura et~al. 1996a,b). At low metallicity, the RSG phase
may be absent; this may also be the case for some binary stars.  As a
first example, we follow the dynamics of the circumstellar medium
around a 29 $M_\odot$ star at the ZAMS, which evolves (at solar
metallicity) through a long-lived RSG stage with prominent
consequences for the evolution of the circumstellar matter. The 29
$M_\odot$ stellar model has steady winds during the main-sequence (MS)
and RSG stages.  For that reason, the CSM evolution during these
stages is computed in one dimension for a homogeneous ISM. For the
calculation of the post-RSG evolution the variables such as the
temperatures and density are extrapolated onto a two dimensional
grid. Note, however, that the calculations start before the RSG phase
has ended in order to be sure that the RSG shell is dynamically
stable.  The total amount of mass lost by the RSG in the wind is
$M_{\rm rsg}=10M_\odot$, with a mass loss rate $\dot{M}_{\rm rsg}=6
\times 10^{-5} M_\odot {\rm yr}^{-1}$ and wind velocity $v_{\rm rsg}
\sim 15$ km s$^{-1}$.

When the fast WR wind $v_w \sim 3500$ km s$^{-1}$ starts blowing, it
sweeps up the RSG wind material into a shell, which we will refer to
as the {\it WR shell}. The properties of this shell turn out to be
very sensitive to the characteristics of the RSG wind. Since the
density of the RSG wind depends on its velocity as $\rho(r)=
\dot{M}_{\rm rsg}/ (4\pi r^2 v_{\rm rsg})$, a low RSG wind velocity
implies a much higher density.  The expansion of the WR shell is
faster for lower RSG wind densities and also for higher wind
velocities. The termination shock of the WR wind is located at a much
smaller radius than that of the hot MS bubble described in the
previous section, so we simply assume free outflow as the outer
boundary condition in order to calculate its dynamical behaviour
(Garc\'{\i}a-Segura et al. 1996a).

\begin{figure}[a]
\centering \noindent {\par\centering
\resizebox*{0.85\columnwidth}{!}{\includegraphics{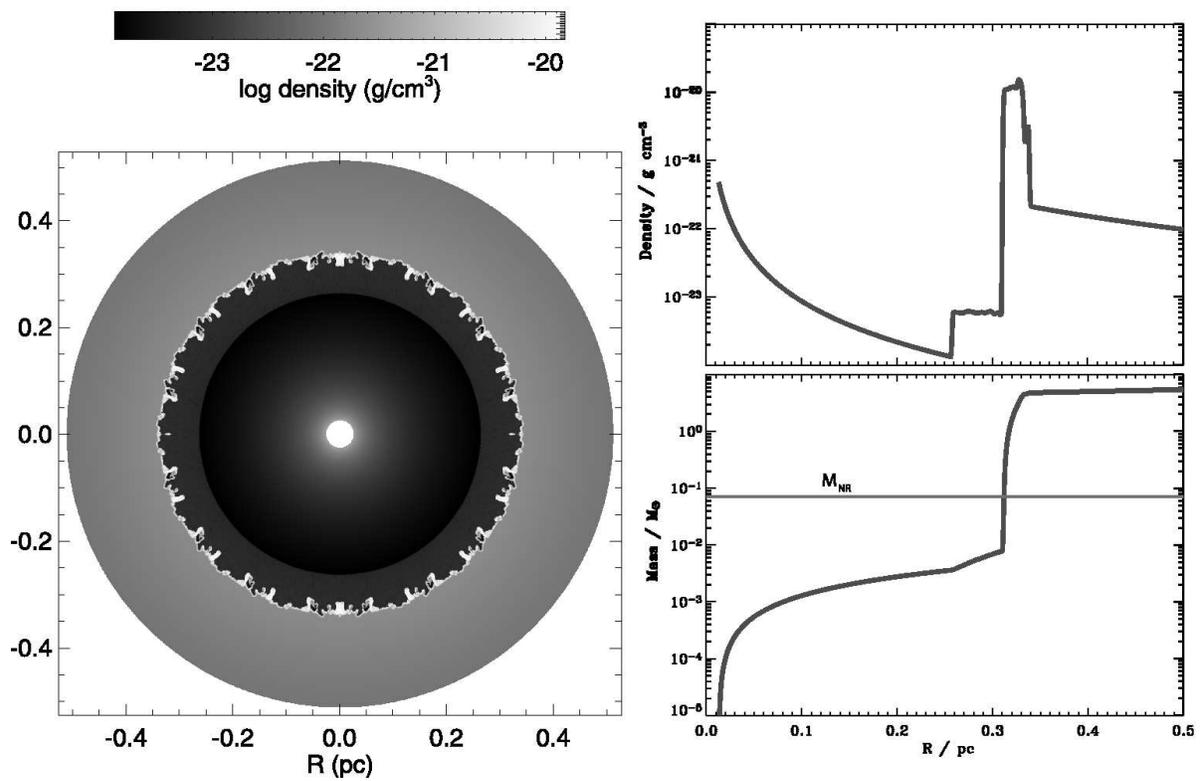}} \par}
\caption{The state of the circumstellar medium around a 29 $M_\odot$
  massive star after 8,000 yr of evolution since the onset of the WR
  phase.  The grid has 800 $\times$ 180 zones, with a radial extent of
  0.5pc and an angular extent of 22.5$^\circ$. The inner-most radial
  zone lies at 0.0125 pc. {\it Left panel:} Logarithm of the
  circumstellar density in units of g cm$^{-3}$. {\it Right Panel}:
  Density and cumulative mass as a function of radius along the polar
  axis.}  
\label{fig:ms}
\end{figure}

During the early stages, the swept-up WR shell is dense enough to be
fully radiative. This ceases to be true, however, after the shell has
extended to a radius of more than 0.1 pc, since its density decreases
so much that its evolution becomes almost adiabatic. Correspondingly,
the shell is initially thin and therefore subject to Vishniac
instabilities (Vishniac 1983); but the increase of shell thickness
soon suppresses the growth of the instability. This instability
operates in dense winds where the cooling is effective enough to
produce radiative terminal shocks. The RSG wind shell is accelerated
by the (less dense) shocked WR wind and is therefore also subject to
Rayleigh-Taylor (RT) and Vishniac instabilities. In this case we have
a radiatively-cooled, thin, accelerated shell and the two
instabilities must be coupled.  The resulting structure after 8,000 yr
of evolution since the onset of the WR phase is summarized in
Fig. \ref{fig:ms}.

It is also possible for a fast WR wind to collide with early ejecta
before it is decelerated by the RSG shell (or a LBV shell depending on
the evolutionary pathway of the massive progenitor).  To illustrate
this, we study the last (unsteady) WR stage of a 60 $M_\odot$
star. Mass loss lowers the stellar mass from 60 $M_\odot$ all the way
down to $\sim 4\;M_\odot$ at the time of the final supernova (or
GRB). The total mass loss includes $\sim 32\;M_\odot$ lost in the MS
and pre-LBV stages, $\sim 8\;M_\odot$ lost in the LBV stage, and $\sim
16\;M_\odot$ in the WR stage. The MS and LBV-WR stages have been
modeled in a self-consistent fashion by Garc\'{\i}a-Segura et
al. (1996b). These stages have important implications for the
circumstellar gas at $r \gg 0.4$ pc. Here, we study the interaction
between the fast-moving WR wind ($v_{\rm wr} \sim 4000$ km s$^{-1}$)
just before the final supernova and the early, slower moving ($v_{\rm
wr} \sim 1500$ km s$^{-1}$), ejecta\footnote{The reader is refer to
Fig. 2 of Garc\'{\i}a-Segura et~al. (1996b) for the stellar mass-loss
rates and wind velocities as a function of time for the 60 $M_\odot$
model.}.

The resultant shell will be pushed outward by the central star wind
and retarded by the early ejected wind, quickly reaching a constant
velocity, but increasing in mass. The resulting expansion law for the
shell follows from the momentum balance between the two components.
The analytical theory of Kwok et al. (1978) can give some insight into
the problem. If $M_{ij}$ is the mass of the resulting shell when it is
at a radial distance $r_{ij}(t)$, then
\begin{equation}
M_{ij}(t)=\int_{r_{j} + v_{j}t}^{r_{ij}(t)} { \dot{M}_j \over v_j}dr +
\int_{r_{ij}(t)}^{r_{i} + v_{i}t}{ \dot{M}_i \over v_i}dr,
\label{eqn:mijw}
\end{equation}
where $v_j < v_i$. If $v_{ij}(t)$ is the velocity of the shell,
then, assuming a completely inelastic collision, the equation of
motion may be written 
\begin{equation}
M_{ij}(t){d\;v_{ij} \over dt} = {\dot{M}_i \over v_i}[v_i - v_{ij}]^2
- {\dot{M}_j \over v_j}[v_j - v_{ij}]^2.
\label{eqn:mijwind}
\end{equation}
Numerical integration of equation (\ref{eqn:mijwind}), with a
substitution for $M_{ij}(t)$ from equation (\ref{eqn:mijw}) gives the
resulting expansion law for the shell. The thickness of the shell
$\Delta r_{ij}$ may be found by requiring its internal pressure to
balance the pressure from the wind.

Such large expansion velocities $\Delta v_w$ produce high temperatures
(see equation \ref{eqn:tshock}) in the post-shock region behind the
outer shock of the merged shell. This high post-shock temperature and
the low density (i.e. low $\dot{M}_{\rm wr}$ and high $v_{\rm wr}$) of
the WR wind result in an almost adiabatic, hot, low-density WR
shell. 

\begin{figure}[a]
\centering \noindent {\par\centering
\resizebox*{0.85\columnwidth}{!}{\includegraphics{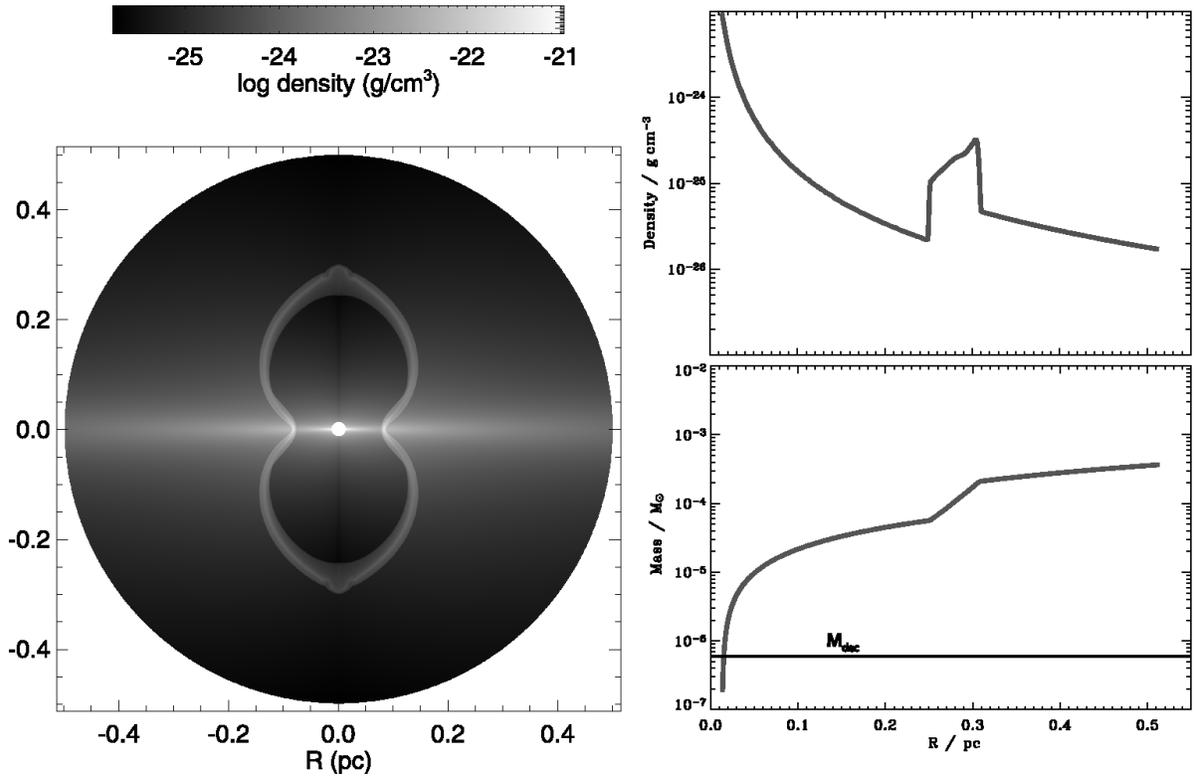}} \par}
\caption{The state of the nearby circumstellar medium around a 60
$M_\odot$ rotating massive star. During the last stages within its WR
lifetime, the wind velocity increases. The fast wind collides with
early ejecta before reaching the LBV shell. The reader is refer to
Fig. 2 of Garc\'{\i}a-Segura et~al. (1996b) for the stellar mass-loss
rates and wind velocities as a function of time. The grid has 800
$\times$ 180 zones, with a radial extent of 0.5pc and an angular
extent of 90$^\circ$. The inner-most radial zone lies at 0.0125
pc. {\it Left panel:} Logarithm of the circumstellar density in units
of g cm$^{-3}$. {\it Right Panel}: Density and cumulative mass as a
function of radius along the polar axis. }
\label{fig:rot}
\end{figure}

Figure \ref{fig:rot} shows the morphology of the smooth WR ring under
the assumption that the central star experiences non-spherical mass
loss close to critical rotation: in other words, a scenario in which a
slower and denser wind is confined to the equatorial plane. To compute
the latitudinal dependence of the wind properties of a star close to
critical rotation ideally requires multi-dimensional models of the
star and its outflowing atmosphere, which are not available. Langer
(1998), however, argued that the stellar flux and the radius might
still vary only weakly from pole to the equator in very luminous
stars. We therefore applied equations similar to those found by
Bjorkman \& Cassinelli (1993) for winds of rotating stars in the limit
of large distance from the star:
\begin{equation}
v_{\infty}(\theta) = \zeta v_{\rm esc} \left(1 - \Omega\, \sin\theta
\right)^{\varphi}\;,
\end{equation}
where we set the parameters defined in Bjorkman \& Cassinelli (1993)
to $\zeta=1$ ,$\varphi=0.35$, $\Omega=v_{\rm rot}/ v_{\rm crit}$, and
$v_{\rm crit}=v_{\rm esc}/\sqrt{2}=[GM_*(1-\kappa)R_*]^{1/2}$, with
$M_*$ and $R_*$ being mass and radius of the star, and $\kappa$
standing for the ratio $L/ L_{\rm Edd}$ of stellar to Eddington
luminosity (Langer et al. 1998). Under the above conditions, the wind
expands more quickly and easily into the lower density wind at the
poles, while stellar rotation concentrates it toward the equatorial
plane, producing a double-lobed structure. In section \S 5 we aim to
examine the interaction of the relativistic blast wave with these more
{\it realistic} density wind profiles.\\

\subsection{The Role of Binarity}

One of the most important questions relating to WR stars is whether
they are all members of binary systems or rather do some truly single
objects exist. Possibly a large fraction of stars are members of
binary or multiple systems but most are sufficiently far from their
companions that their evolution proceeds essentially as if they were
single stars. However there is an important minority of stars which
are close enough that their evolution is dramatically changed by the
presence of a companion. It is extremely unlikely that the progenitors
of GRBs are just very massive, single WR stars. Special circumstances
are almost certainly needed. The most promising of these is rotation:
a rapidly rotating core is the essential ingredient in the "collapsar"
model for GRBs (Woosley 1993; MacFadyen \& Woosley 1999). Massive
stars are generally rapid rotators on the main sequence. However,
there are many well-established mechanisms by which they can lose
their angular momentum during their evolution. Thus, it is not at all
clear whether the cores of massive single stars will ever be rotating
rapidly at the time of explosion (e.g. Heger et al. 2005).  The
simplest way to generate the required angular momentum is for the core
to reside in a tight binary and be in corotation with the
binary. Izzard et al. (2004), using detailed binary population
calculations, concluded that there are enough binaries where tidal
locking could account for the observed GRB rates. Most companions,
being low-mass MS stars, have small mass-loss rates and are thus
unlikely to enhance significantly the density around the GRB
progenitor.  The density structure seen around WR 147 (Contreras \&
Rodriguez 1999) could be a hint of what exists in some GRB
progenitors. Overall, we do not expect binarity to significant alter
the circumstellar gas density profiles along the rotation axis of the
collapsing stellar core. There is, however, a small, interesting
subclass of binaries in which the companion star may be massive (see
e.g. Izzard et al. 2004). In this case, the WR star mass loss may also
be influenced by the winds of the close binary companion. Detailed
computations are needed to show the magnitude of the effect.

The requirement that the total core angular momentum exceed the
maximum angular momentum of a Kerr black hole places interesting
limits on the binary period (Izzard et al. 2004; Podsiadlowski et
al. 2004). We make the reasonable assumption that the binary is
circular, and model the core as an $n=3$ polytrope. then the binary
period must be smaller than $P_{\rm orb} \sim 4 (M_{\rm core}/2
M_\sun)^{-1}(R_{\rm core}/10^{10}\;{\rm cm})^{2}$ h. This orbit could
be tight enough that the core may in fact have been stripped of its
helium in a common envelope to form a CO core. Alternatively, it may
sometimes happen that the core of a very massive star retains the
required angular momentum as its outer hydrogen layers are blown off
in a stellar wind. Indeed, Brown \& Bethe (1994) have noted that the
cores of stars more massive than $\sim 20 M_\sun$ will undergo a
delayed collapse to form a black hole if the nuclear equation of state
is soft. Prompt formation of a black hole introduces a mechanism for
failure of a core collapse supernova, if success of the shock depends
on delayed neutrino heating. The corresponding rate of Kerr hole
formation depends, of course, on the physics of angular momentum
transport inside the progenitor star. The required spin could also be
generated when the core mergers with a binary companion during common
envelope phase.  So perhaps one important distinction between a GRB
and an ordinary supernovae is whether a black hole or a neutron star
is formed in the aftermath.  However, not all black hole formation
events can lead to a GRB: if the minimum mass of a single star that
leads to the formation of a black hole is as low as $25 M_\sun$, this
would overproduce GRB by a large factor (see Izzard et al. 2004).

\section{The Afterglow Appearance}
The interaction of a relativistic blast wave with the surrounding
medium is described by the adiabatic Blandford \& McKee (1976;
hereafter BM) self-similar solution. The scaling laws that are
appropriate for the burst interaction with a medium with particle
density $n \propto r^{-s}$ have been described by M\'esz\'aros et
al. (1998), Chevalier \& Li 2000, Panaitescu \& Kumar (2000), Wang et
al. (2000), Ramirez-Ruiz et al. (2001), Dai \& Lu (2002), and Dai \&
Wu (2003).

For an adiabatic ultra-relativistic blast wave, the (isotropic
equivalent) total energy is
\begin{equation}
E= {8\pi A \Gamma^2 r^{3-s}c^2 \over 17 - 4s},
\end{equation}
where $\Gamma$ is the bulk Lorentz factor of the shock front and $r$
is the observed radius near the line of sight (BM). We assume the
burst to be collimated with an initial half-angle $\theta$ larger than
20$^{\circ}$, and that lateral expansion is negligible during the
relativistic phase. A distant observer receives a photon emitted along
the line of sight towards the fireball center at a time (Chevalier \&
Li 2000)
\begin{equation}
t={r \over 4(4-s)\Gamma^2c}, 
\end{equation}
and so
\begin{equation}
r = \left[{(4-s)(17-4s)Et \over 2\pi Ac }\right]^{1\over(4-s)}.
\end{equation}
Before the collision with the high-density shell, the shock front is
expected to propagate through an $n(r)=Ar^{-2}$ wind, where
\begin{equation}
A= 3 \times 10^{35} \dot{M}_{{\rm wr},-5} v_{{\rm wr},3}^{-1}\; {\rm
cm}^{-1}
\end{equation}
for a spherically symmetric wind ejected at a constant speed.  The
relativistic expansion is gradually slowed down until the shock
front encounters the wind density discontinuity at a radius $r_{\rm
sh}$. A transition in the afterglow is then observed
\begin{equation}
t_{\rm sh} \approx E_{52}^{-1} r_{{\rm sh},17}^2 A_{35}\;{\rm days}
\end{equation}
after the burst. This of course only true for $r_{\rm sh} < r_{NR}$
(see equation \ref{eqn:mnr}). This encounter produces two new shock
waves: a forward shock that moves into the wind thin shell
discontinuity and a reverse shock that propagates into the
relativistic ejecta. In the section that follows we examine the state
of these shocks under various assumptions regarding the density
discontinuity.

\subsection{Shock Conditions}
\begin{figure}[a]
\centering \noindent {\par\centering
\resizebox*{0.85\columnwidth}{!}{\includegraphics{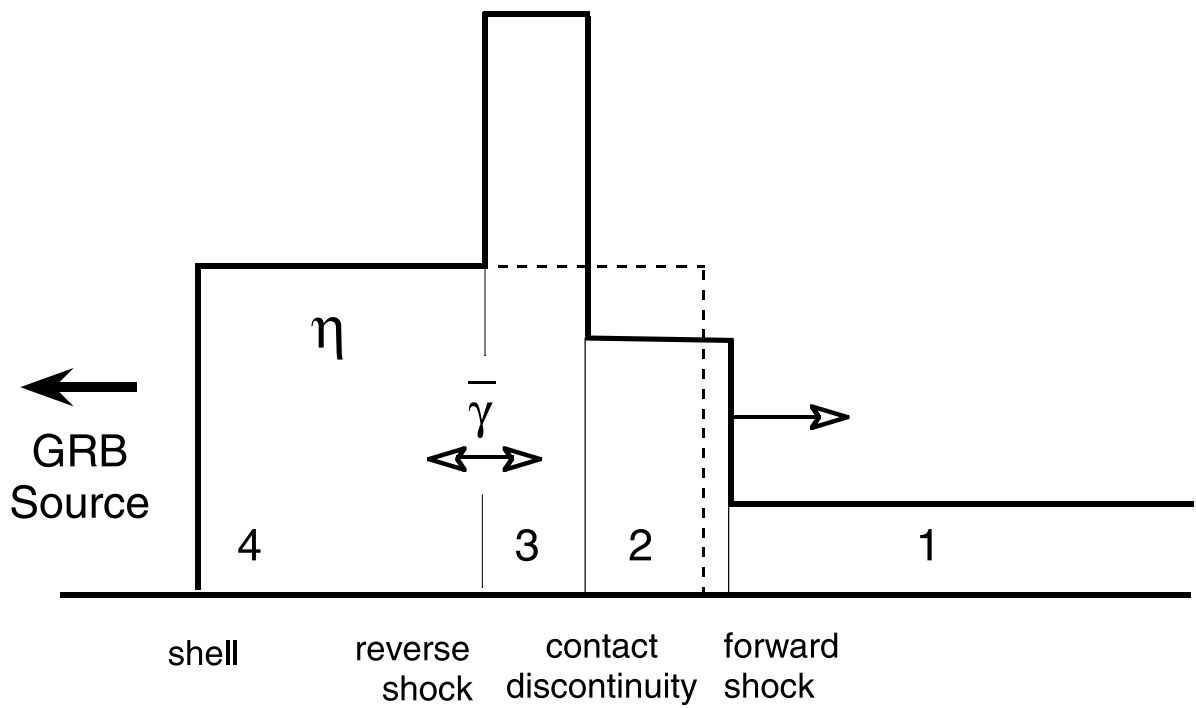}}
\par}
\caption{Diagram illustrating a relativistic shock system. Basic
system consisting of a shocked fluid encountering matter at
rest. Quantities for the system are the unshocked ejecta, the reverse
and forward shocks, and the external matter at rest. The dashed line
schematically shows the properties of the relativistic ejecta in the
absence of an external medium.}
\label{fig:diag}
\end{figure}

Consider a relativistic shell with a Lorentz factor $\eta$ moving into
the cold circumburst medium (CM). The interaction is described by two
shocks: a forward shock propagating into the CM and a reverse shock
propagating into the ejecta. There are four regions separated by the
two shocks: the CM (1), the shocked CM (2), the shocked ejecta (3) and
the unshocked ejecta (4). The CM is at rest relative to the observer
(Fig. \ref{fig:diag}).  Velocities $\beta_i$, and their corresponding
Lorentz factors $\gamma_i$, distances and time are measured relative
to this frame. Thermodynamic quantities $n_i$, $p_i$ and $e_i$
(particle number density, pressure, and internal energy density) are
measured in the fluids' rest frames. The ISM and the unshocked shell
are cold and, therefore, $e_1=e_2=0$. The shocked material is
extremely hot and, therefore, $p_2=e_2/3$ and $p_3=e_3/3$ (Sari \&
Piran 1995).

For $\eta \equiv \gamma_4 \gg 1$ the equations governing the shocks
are (BM)
\begin{equation}
{e_2 \over n_2m_pc^2}\;=\;\gamma_2\;-\;1\; \approx \;\gamma_2;\;\;\; {n_2 \over n_1}\;=\;4 \gamma_2\;+\;3 \;\approx 4\gamma_2
\label{eqn:shock1}
\end{equation}
\begin{equation}
{e_3 \over n_3m_pc^2}\;=\;\overline{\gamma}_3\;-\;1;\;\;\; {n_3 \over
n_4}\;=\;4\overline{\gamma}_3 \;+\;3.
\label{eqn:shock2}
\end{equation}
The approximations in equation (\ref{eqn:shock1}) use only the fact
that $\gamma_4 \gg 1$ and therefore $\gamma_2 \gg 1$. No assumption is
made about $\overline{\gamma}_3$, the Lorentz factor of the motion of
the shocked material in region (3) relative to the unshocked shell in
region (4).

Equality of pressures and velocities along the contact discontinuity yields 
\begin{equation}
e_2=\;e_3;\;\;\overline{\gamma}_3 \approx {1 \over 2}\left({\gamma_4
\over \gamma_2}+{\gamma_2 \over \gamma_4}\right).
\label{eqn:shock3}
\end{equation}
The solution for $\gamma_2$ depends only on two parameters, $\eta$ and
$\psi \equiv n_4/n_1$. The energy, pressure, and density also depend
linearly on a third parameter, the external density $n_1$ (Sari \&
Piran 1995).

There are two simple limits of equations
(\ref{eqn:shock1})-(\ref{eqn:shock3}) in which the reverse shock is
either Newtonian or ultra relativistic (the forward shock is always
ultra relativistic if $\eta \gg 1$ and $\psi > \eta^{-2}$). If $\eta^2
\gg \psi$, the reverse shock is ultra relativistic ($
\overline{\gamma}_3 \gg1$):
\begin{equation}
\overline{\gamma}_3\; =\; {\sqrt{\eta} \over
\sqrt{2}\psi^{1/4}};\;\;\;\gamma_2\;=\;{\gamma_3\;=\;{\sqrt{\eta}\psi^{1/4}
\over \sqrt{2}}}.
\label{eqn:relshock}
\end{equation}
In this case almost all of the initial kinetic energy is converted by
the shocks into internal energy ($\eta \gg \gamma_3$). The process
therefore is over after a single passage of the reverse shock through
the ejecta (Sari \& Piran 1995).

If $\psi \gg \eta^2$ the reverse shock is Newtonian
$(\overline{\gamma}_3\; -\; 1 \ll 1)$ and
\begin{equation}
\overline{\gamma}_3-1\;\approxeq\;{4\eta^2\psi^{-1} \over 7}\equiv 2\epsilon\; \ll 1;\;\;\; \gamma_2=\gamma_3=\eta(1-\sqrt{\epsilon}).
\end{equation}
The reverse shock converts only a fraction $\gamma/\sqrt{\psi}\ll 1$
of the kinetic energy into internal energy. It is too weak to slow
down the ejecta effectively, and most of the initial energy is still
kinetic energy when this shock reaches the inner edge of the
ejecta. At this stage a rarefraction wave begins to propagate toward
the contact discontinuity. This wave propagates at the speed of sound
$\sqrt{4p_3/(3n_3m_p)}$ and it reaches the contact discontinuity at
$t_r=(3\sqrt{7}/4)\Delta \eta\sqrt{\psi}/c$, which is of the same
order of magnitude as the shock crossing time. Here $\Delta$ is the
width (in the observer's frame) of the relativistic shell. The
rarefraction wave is then reflected from the contact discontinuity and
a second, weaker, shock wave forms. A quasi-steady state slowing down
solution forms after a few crossing like this (Sari \& Piran
1995). Using momentum conservation, the total slowing down time can be
estimated by $\sim \gamma \Delta n_4m_p c/p_2$. During this time the
forward shock collects a fraction $\gamma^{-1}$ of the shell's rest
mass. In contrary to the relativistic case, there are two relevant
timescales now: the rarefraction or shock crossing time and the total
slowing down timescale.

In any realistic situation the CM is probably inhomogeneous, as in the
stellar models described in the previous section. Consider a density
jump by a factor $\alpha$ over a distance $\lambda$. The forward shock
propagates into the CM with a density $n_1$ as before, and when it
reaches the position where the CM density is $\alpha n_1$ a new shock
wave is reflected. This shock is reflected again off the shell. As
discussed above, the reflections time is about $t_\alpha \sim \lambda/
(4 c \alpha^{1/2})$, and after these reflections, the Lorentz factor
and hydrodynamical properties of the system are as if the CM were
homogeneous with a density $\alpha n_1$. If the reverse shock converts
only a small fraction of the kinetic energy into thermal energy, then
the forward shock expands almost at the velocity of the previous blast
wave. In the relativistic case, however, considerable deceleration may
occur before $t_\alpha$ unless $t_\Delta \sim \eta \sqrt{\psi}
\Delta/c \gtrsim t_\alpha$.

The peak of synchrotron emission occurs at frequency $\nu_m \approx
q_e \gamma_e^2 B/(m_ec)$ and the frequency integrated emissivity is
given by $\epsilon \approx \sigma_T c n_e \gamma_e^2 B^2/8\pi$; where
$q_e$, $m_e$, and $\gamma_e$ are electron charge, mass, and thermal
Lorentz factor, respectively, $B^2/8\pi=\varepsilon_{B} n_e \gamma_e
m_e $ is the magnetic field in the fluid rest frame (assumed to be
amplified up to a fraction $\varepsilon_{B}$ by the processes in the
shocked region and not determined directly by the field in the WR
star), $n_e$ is the electron number density, and $\sigma_T$ is the
Thomson cross-section. The emission at low frequencies ($\nu \ll
\nu_{\rm m}$) scales as $\nu^{1/3}$ and at high frequencies ($\nu \gg
\nu_{\rm m}$) it scales as $\nu^{-(p-1)/2}$, where $p\approx 2.5$ is
the power-law index for the energy distribution of electrons. The
ratio of the peak synchrotron frequencies, before and after the
forward shock has traversed a density jump of a factor $\alpha$ over a
distance $\lambda$, is
\begin{equation}
{\nu_{{\rm m,\alpha}} \over \nu_{\rm m}} = \left({e_{2,\alpha} \over
e_2}\right)^{5/2}\left[{n_2 \over n_{2,\alpha}}\right]^{2} \approx
\alpha^{-1/4}.
\end{equation}
Therefore, we expect this emission to be a significant contribution to
the long-wavelength flux\footnote{The above estimate assumes that the
reverse shock is initially relativistic, which is likely to be the
case when the density contrast of the density discontinuity is high
(Dai \& Lu 2002).}. The ratio of the observed flux after the forward
shock has transversed through the density discontinuity, and the flux
in the absence of the collision at a frequency much greater than the
peak of the emission is given by
\begin{equation}
{f_{2,\alpha} \over f_2} =\gamma^{'}_{\alpha}\left({e_{2,\alpha} \over
e_2}\right)^{1/2}\left({\nu_{\rm m,\alpha} \over \nu_{\rm m}}
\right)^{p-1\over 2} \approx \alpha^{-{1 \over 8}(p-1)}
\end{equation}
where $\gamma^{'}_{\alpha}=\gamma_{2,\alpha}/\gamma_2$ is a factor by
which the Lorentz factor of the outer shell decreases as a result of
the collision. The dependence of the observed flux on
$\gamma^{'}_{\alpha}$ can be more rapid than the linear function
considered above, depending on the temporal profile of the
deceleration of the ejecta (see e.g. \S 5.2). The decrease in the
observed emission from the forward shock due to the density jump at a
frequency much greater than the peak of the emission is therefore
given by $\alpha^{-{1 \over 8}(p-1)}\approx \alpha^{-{3 \over 16}}$,
neglecting the enhanced energy losses from the shocks that may arise
during the collision. This is clearly not the case at frequencies
either below $\nu_{\rm m}$ or for which there is significant emission
from the reverse shock (typically a factor of $\eta$ lower in
frequency).  At these frequencies, the observed flux is expected to
increase.

\subsection{Afterglow Lightcurves}
We now generalize the above results to a spatially varying external
density. The afterglow modeling used here is similar to that described
in Salmonson (2003). The jet deceleration is calculated from the mass
and energy conservation equations. The lateral spreading of the jet is
neglected.  The calculation of radiative losses includes synchrotron
emission, and the synchrotron spectrum is taken to be a piecewise
power law with the usual self-absorption, cooling, and injection break
frequencies, calculated from the cooled electron distribution and
magnetic field. The observed flux is obtained by integrating the jet
emission over the equal arrival time surface.

Strong temporal variations compared to the canonical power-law decay
can be produced by changes in the circumburst density or by energy
variations (e.g. Meszaros et al. 1998). For example, as shown in
Ramirez-Ruiz, Merloni \& Rees (2001), a shock wave that has been
slowed by the surrounding medium could be caught up by subsequent
shocks, increasing the shock wave energy. Alternatively, the shock
wave's energy may have varied as the result of encountering an
external medium of variable density. While here we concentrate on the
case where the dominant variations are in density, the two
alternatives can be distinguished by exploiting the fact that the flux
at $\nu> \nu _{c}$ is insensitive to variations in the ambient
density, whereas below the cooling frequency $F_{\nu } (\nu <\nu_{c})
\propto n^{1/2}$ (Nakar et al. 2003; Heyl \& Perna 2003). As the blast
wave expands into the ambient matter, its kinetic energy is used to
shock and heat the matter. Deceleration due to this starts in earnest
when about half the initial energy is transferred to the shocked
matter. The characteristic mass $M_{\rm dec}$ where this takes place
is given by equation (\ref{eqn:mdec}). This phase ends when so much
mass shares the energy that the Lorentz factor drops to 1. This occurs
at a mass scale $M_{NR}$ (see equation \ref{eqn:mnr}).

\begin{figure}[a]
\centering \noindent {\par\centering
\resizebox*{0.85\columnwidth}{!}{\includegraphics{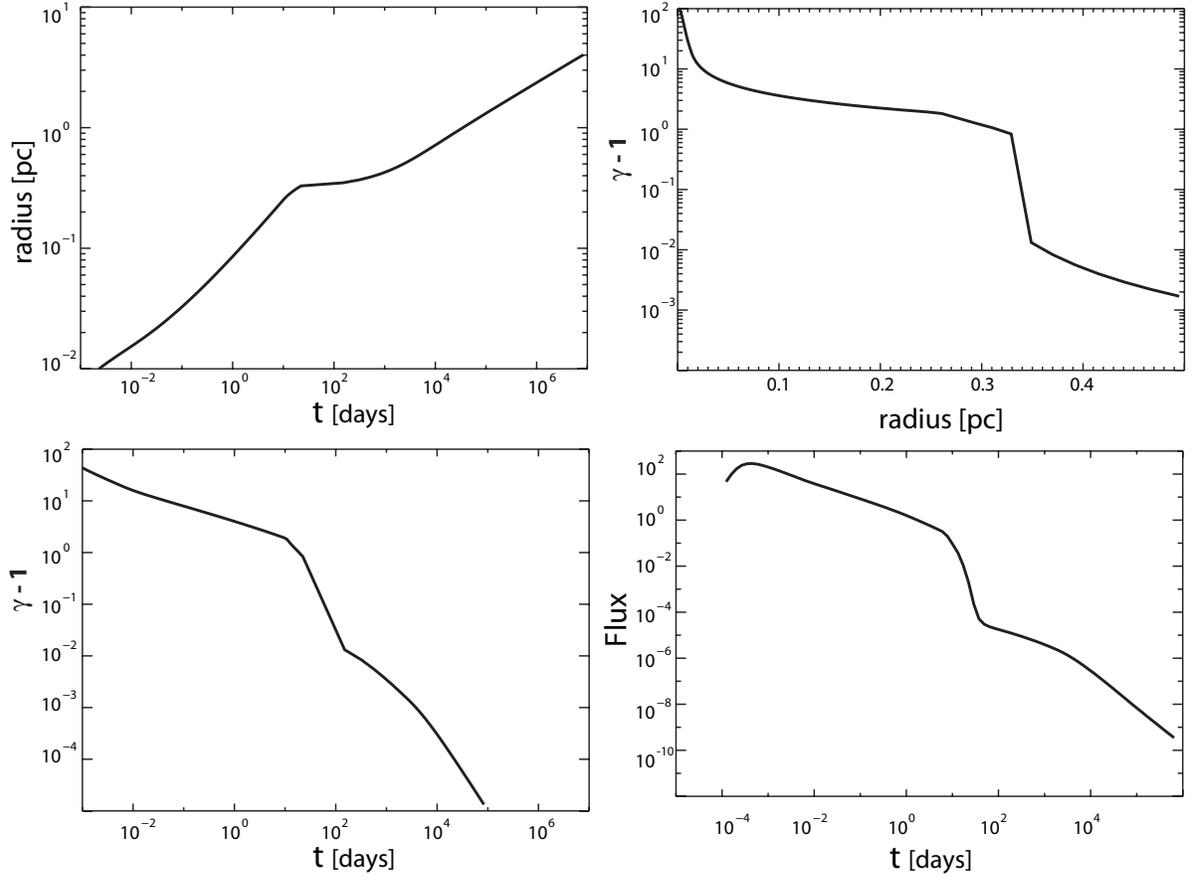}} \par}
\caption{The effect of the impact of a relativistic jet with the wind
density discontinuity on the R-band afterglow.  The shock front
expands within the $s=2$ stellar wind until it reaches the
high-density shell at a distance $r_{\rm sh} \approx 0.2$ pc (see
Fig. \ref{fig:ms}). The shock transit of the massive shell causes a
rapid decline in Lorentz factor and a corresponding decline in flux
(at a fixed frequency). The remaining evolution of the shock is
non-relativistic.  At the time of the collision the relativistic shell
Lorentz factor is $\gamma \sim 4$ for $E= 5 \times 10^{53}$ ergs.  In
this simulation the jet opening angle is $\theta_j=5^\circ$ and it is
viewed at $\theta_{\rm obs} = 3^\circ$ from the jet axis. The
afterglow emission is calculated in the adiabatic regime.  The
collision model takes into account the fireball geometrical curvature
when calculating the photon arrival time and relativistic boosting.}
\label{fig:Flux}
\end{figure}

In the case depicted in Fig. \ref{fig:ms}, for example, the stellar
wind is not dense enough to slow down the ejecta to non-relativistic
speeds before reaching $r_{\rm sh}$, so that we expect part of the
blast wave evolution as we see it to take place outside the $1/r^2$
density distribution. In a rarefied wind such as that illustrated in
Fig. \ref{fig:rot}, the shock front will expand within the stellar
wind until it reaches the density discontinuity at about 0.2 pc
without significant deceleration. Over the typical time of observation
of a GRB afterglow, the impact of the relativistic ejecta with the
density enhancement will produce a clear feature in the observed
emission. Fig. \ref{fig:Flux} and its caption summarize the predicted
R-band afterglow lightcurve from hours until about a few years after
the burst. One can see sharp break in the flux decay curve that
coincides with a precipitous drop in Lorentz factor when the afterglow
shock meets the shell.  In this model, the isotropic energy of the
ejecta is $5 \times 10^{53}$ erg. The characteristic synchrotron
frequency is lower than that in the absence of the collision. This
effect is responsible for the decrease in flux (at a fixed frequency)
seen at about $\sim 10$ days. Further observable transitions are
produced as the blast-wave plows deeper into the shocked-wind
discontinuity. Jet effects are expected to become important at a time
\begin{equation}
t_{j} \approx 5 (\theta_j/0.08)^{4} (E/10^{53} {\rm erg})(A/10^{35}
{\rm cm}^{-1})^{-1} {\rm days},
\end{equation}
just before the impact takes place. This makes the jet-break time not
easily identifiable.

\begin{figure}[a]
\centering \noindent {\par\centering
\resizebox*{0.85\columnwidth}{!}{\includegraphics{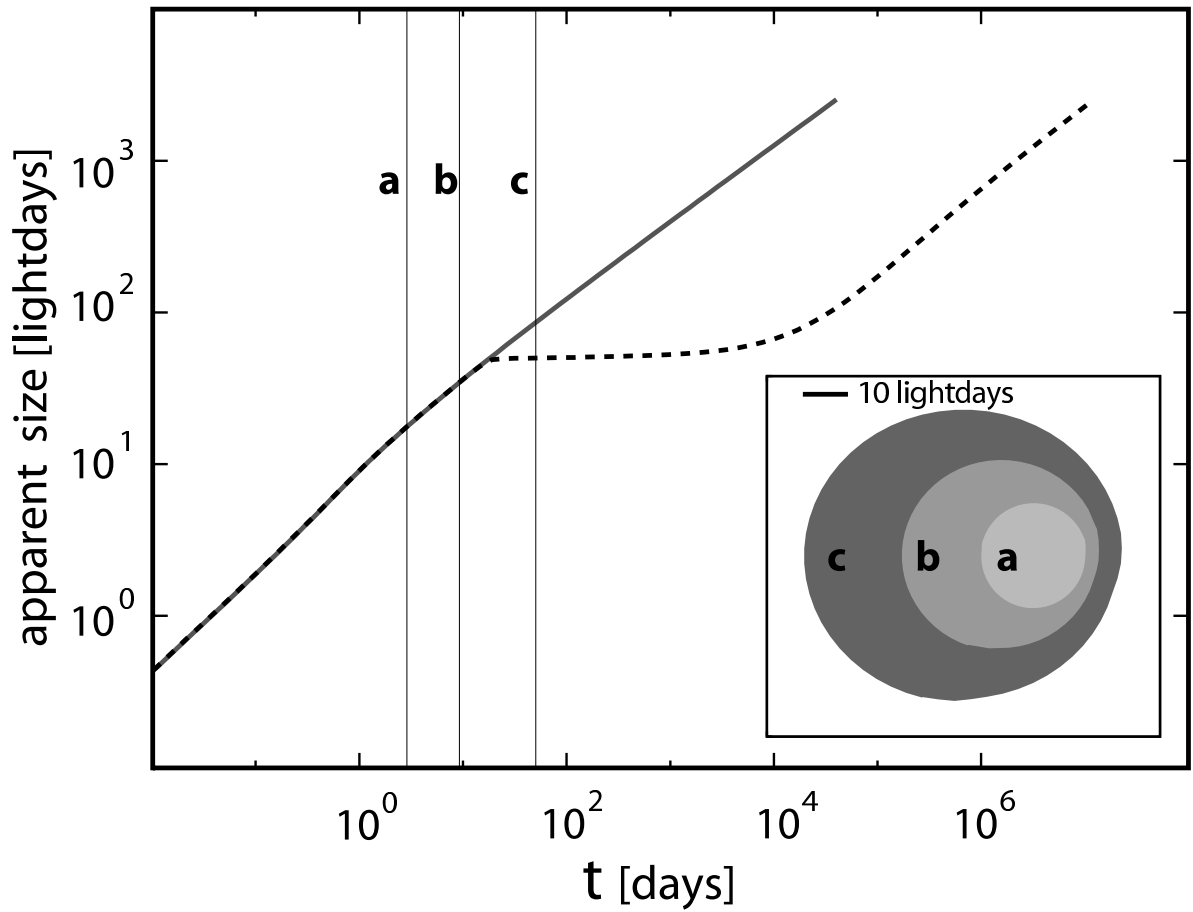}}
\par}
\caption{Evolution of the afterglow source size as a function of time
for a sharp-edged, homogeneous jet seen at $\theta_{\rm obs}=0.6
\theta_j=3^\circ$. This one dimensional, hydrodynamic model takes on
the density profile of the stellar wind in the polar direction.  The
jet deceleration is calculated from the mass and energy conservation
equations. The lateral spreading of the jet is neglected.  The effect
of the impact of a relativistic jet with the wind density
discontinuity (Fig. \ref{fig:ms}) can be seen by comparing the solid
and dashed curves.  For simplicity, the density profile seen in Fig.
\ref{fig:ms} has been extrapolated to large radii. Substantial
deviations in the density profile are, however, expected to occur with
the inclusion of the wind termination shock region (see e.g. Fig. 2 in
Garc\'{\i}a-Segura et~al. 1996a). The inset panel shows the afterglow
source diameter. The x- and y-axes are shown in their true scale and
measured in light-days.}
\label{fig:R_perp}
\end{figure}

The evolution of the apparent source size is shown in
Fig. \ref{fig:R_perp}. The afterglow image is limited to a circle on
the sky, whose size grows as $t^{(5-s)/2(4-s)}$ (Granot \& Loeb 2001).
The assumption of a spherical flow may also serve as an adequate
description of a jetted flow, at sufficiently early times before the
jet break time, $t_j$, when the Lorentz factor of the flow drops below
the inverse of the jet opening angle (Rhoads 1997). In the case of a
jet, the image at $t \gtrsim t_j$ is expected to be different than in
the spherical case (Oren, Nakar \& Piran 2004; Granot, Ramirez-Ruiz \&
Loeb 2004) and will no longer be circular for observers who are
situated away from the jet axis (Fig. \ref{fig:R_perp}). The image
size depends decisively on the density profile of the ambient medium
into which the GRB fireball propagates.  The effects of the density
discontinuity on the size of the afterglow image can be seen by
comparing the solid and dashed curves in Fig. \ref{fig:R_perp}. This
is not altogether surprising: the ejected material is decelerated by
the external medium at a smaller radius than it would be in the
absence of the density discontinuity. The calculations above
demonstrate how the measurable properties of such afterglows depend on
the nature of the progenitor star and the medium around it.

\section{Clump Formation and Afterglow Variability}
The smoothness or clumpiness of the swept-up shell may be responsible
for small-scale variability observed in many gamma-ray burst
afterglows. As illustrated in Fig. 4, the required degree of clumping
in the Wolf-Rayet star wind itself does not seem plausible. On the
other hand, as shown in Fig. 3, there is the possibility of clump
development from the swept up RSG wind. As stated above, the shocked
RSG wind is accelerated by the (less dense) shocked WR wind and is
therefore subject to RT instabilities. The shell is initially thin and
is therefore also subject to Vishniac instabilities. The two
instabilities must then be coupled. The coupling mechanism appears to
be a displacement of gas towards the interior of the bubble at the
contact surface by the RT instability, which induces the formation of
a valley in the outer shock, driving what is fundamentally a Vishniac
instability. Garc\'{\i}a-Segura et al. (1996b) found that clump formation
is not efficient for the case of pure RT instabilities in thick shells
and concluded that a necessary condition for clump formation to occur
is that a shell be thin enough to allow Vishniac-like instabilities to
drive transverse motions to form the clumps. Many RT fingers will then
warp and break, forming knots.

Once formed, the knots tend to dissipate. Several process tend to
destroy the inhomogeneities embedded in a wind-blown bubble, namely
thermal evaporation (Cowie \& McKee 1977), ablation (Hartquist et
al. 1986) and photo-evaporation (McKee 1986).  If not inhibited by the
presence of a magnetic field, heat conduction may deplete the
clumps. The thermal evaporation time is 
\begin{equation}
t_{\rm ev} \propto {n_{\rm c} T_{\rm sw} \Delta r \over n_{\rm sw}},
\end{equation}
where $T_{\rm sw}$ is the temperature of the hot shocked WR wind,
$\Delta r$ and $n_{\rm c}$ are, respectively, the radii and densities
of the clumps, and $n_{\rm sw}$ is the shocked wind density. For
$n_{\rm c}\ge 10 n_{\rm sw}$, $T_{\rm sw} \sim 4 \times 10^7$ K and
$\Delta r \leq 0.01$ pc one has $t_{\rm ev} \leq 10^4$ yr, so that, if
conduction is unimpeded, the lifetime of the clumps may be shorter
than the nebula lifetime. Erosion via ablation is far less effective,
and the clumps survive if, as expected (see e.g. Hartquist et
al. 1986), heat conduction is inefficient. The ablation time for a
knot is
\begin{equation}
t_{\rm ab} \propto {n_{\rm c} \Delta r \over n_{\rm sw} v_{\rm rsg}}.
\end{equation}
We estimate $t_{\rm ab}\ge 10^2 t_{\rm ev}$. The characteristic
photo evaporation time of a clump at a distance $r$ from a star
emitting $S$ ionizing photons per second is 
\begin{equation}
t_{\rm pe}(r) \propto n_{\rm c} {\Delta r^{3/2} \over S^{1/2}} r.
\end{equation}
For the clump parameters considered above and $S=10^{49}$ s$^{-1}$ one
has $t_{\rm pe}\sim 10^4\;r_{\rm pc}$ yr. Small neutral condensations
are thus likely to be photo-evaporated while larger ones survive.

The jet will then encounter clumps -- if they are not efficiently
destroyed -- with large variations in density. Presumably the clump is
symmetric such that
\begin{equation}
\Delta r_{\parallel}=\Delta r_{\perp}=\Delta r_{\rm clump}.
\end{equation}
Whether or not the ambient object collapses onto the shell is a key
distinction that must be made in order to understand how the clump
determines the time structure.  We assume here the case of a
"collapsible" object (such as a wind clump). The contribution to the
peak duration from the time the jet takes to move through the clump
$\Delta r_{\rm clump}/(2 \gamma^2 c)$, is negligible compared with the
time the shell takes to engage the perpendicular size of the
object. This engagement time is caused by the curvature of the
jet. The curvature of the expanding jet prevents it from engaging the
cloud instantaneously.  Rather, the portion of the jet at $\theta \sim
\gamma^{-1}$ requires a time $r(1-\cos\theta)/v$ longer to reach the
cloud. At an angle $\theta$ from the line of sight, the time to engage
the object is $\theta \Delta r_{\rm clump}/2c$. At a typical angle of
$\theta \sim \gamma^{-1}$,
\begin{equation}
\Delta t ={2 \Delta r_{\rm clump} \over c\gamma}.
\end{equation}
The alternative is that the ambient source does not collapse, but
produces photons on a scale of $\Delta r_{\rm clump}$ (unlikely).  The
timescale in this case is determined from the light travel time across
the overlap of the shell thickness and the ambient source thickness.
If the clumps where distributed as a power-law of clump sizes, one
might expect a power law distribution of observable peak widths. The
cumulative effect of many small-scale density perturbations will tend
to average out during the expansion history of the jet, and,
therefore, the resultant afterglow profile should be relatively
smooth. An upper limit on $\Delta r_{\rm clump}$ is set by observed
afterglow variability timescales (i.e. $\Delta t_{\rm obs}$), where
\begin{equation}
c\Delta t_{\rm obs} < \Delta r_{\rm clump} < {c\Delta t_{\rm obs}
\gamma \over 2}.
\end{equation}
Short time scale of afterglow oscillations provide interesting upper
limits of $\Delta r_{\rm clump} \leq 10 (\gamma/10)$ AU on the size of
the clumps around GRB 011211 (Jakobsson et al. 2004) and GRB 030329
(Lipkin et al. 2004). These limits are lower than the fluctuation
amplitudes seen on similar scales in the local interstellar medium
(Wang \& Loeb 1999), though they may reflect the length scale of
comet-like clumps produced in ring nebulae surrounding massive stars.

\section{Discussion}
It is evident from the above discussion that the environment of a
massive star at the time of its death is a very rich one. Even in the
simplest case of a wind whose properties do not vary over the life of
the star, complex behaviour with multiple possible transitions in the
observable part of the afterglow lifetime may be seen.  The eventual
resulting afterglow lightcurve depends fairly strongly on the
properties of the system, especially the mass-loss rate of the star
and the ambient density. This has a good and a bad side. On the
negative side, it implies that one can not be too specific about the
times at which we expect to see transitions in the observed
emission. On the positive side, if and when we do see these
transitions, they can be fairly constraining on the properties of the
stellar progenitor.

Considering the radial range relevant to GRBs, the absence of the
expected $1/r^{2}$ density structure in many bursts is not
surprising. On the other hand, the reason for low densities remains
unclear. A wind termination shock may resolve this matter although
very special conditions are necessary to bring this about.  Because
the density in a shocked wind is higher than that in a free wind at
the same radial distance, the low-density requirement is not
alleviated by appealing to a shocked WR wind. A low stellar mass wind
loss, a faster wind velocity or a low metallicity may help, although
it is not clear whether realistic assumptions can provide the required
low densities (Wijers 2001; Chevalier et al. 2004). The following
point should be emphasized here. The fireball model used to infer the
circumburst density is highly simplified. For example, the wind
density is assumed to follow a pure $1/r^2$ law and to be free of
inhomogeneities, the expanding jet surface is assumed uniform with no
internal (density, velocity) gradients and the fraction of the
explosion energy in the post shock magnetic field is assumed to be
constant (Panaitescu \& Kumar 2002). We do not know whether a
description of the afterglow data is possible with a wide variety of
underlying assumptions and whether that would substantially change the
parameters inferred. Still, the low densities inferred by afterglow
observations are thus problematic for the collapsar model.

The task of finding useful progenitor diagnostics is simplified if the
pre-burst evolution leads to a significantly enhanced gas density in
the immediate neighborhood of the burst. The detection of spectral
signatures associated with the GRB environment would provide important
clues about the triggering mechanism and the progenitor. A special
case is that of GRB 021004, where lines of highly ionized species,
blueshifted relative to the host galaxy, have been attributed to a
Wolf-Rayet stellar wind (Mirabal et al. 2003; Schaefer et al. 2003).
Stars interact with the surrounding interstellar medium, both through
their ionizing radiation and through mass, momentum and energy
transfer in their winds. Mass loss leads to recycling of matter into
the interstellar medium, often with chemical enrichment.  Mass loss is
a significant effect in the evolution of massive stars; in particular,
for stars that enter WR stages (e.g. Ramirez-Ruiz et al. 2001). WR
stars follow all or part of the sequence WNL, WNE, WC and WO, which
corresponds to a progression in the exposure of nuclear products (CNO
equilibrium with H present, CNO equilibrium without H, early
visibility of the products of the 3$\alpha$ reaction and then a
growing (C+O)/He ratio respectively).  Mass loss drastically
influences stellar yields. At low $Z$ there is a high production of
$\alpha$-nuclei, while at higher $Z$ most of the He and C produced is
ejected in stellar winds and escapes further nuclear processing.This
characteristic offers a direct observational test of which stars are
likely to produce a GRB.

Finally, The total energy observed in $\gamma$-rays from GRBs whose
redshift has been determined is diverse (e.g. Soderberg et
al. 2004). One appealing aspect of a massive star progenitor is that
the great variety of stellar parameters can probably explain this
diversity. Given the need for a large helium core mass in progenitors,
the burst formation may be favored not only by rapid rotation but
also by low metallicity. Larger mass helium cores might have more
energetic jets, but it is unclear whether they can be expected to be
accelerated to large Lorentz factors (MacFadyen, Woosley \& Heger
2001; Ramirez-Ruiz, Celotti \& Rees 2002). Many massive stars may
produce supernovae by forming neutron stars in spherically symmetric
explosions, but some may fail neutrino energy deposition, forming a
black hole in the center of the star and possibly a GRB. One expects
various outcomes ranging from GRBs with large energies and durations,
to asymmetric, energetic supernovae with weak GRBs (Totani 2003;
Granot \& Ramirez-Ruiz 2004). The medium surrounding a GRB would
provide a natural test to distinguish between different stellar
explosions.

\acknowledgments We thank M.~J. Rees, J. Granot, R. Chevalier,
S. Woosley and N. Langer for helpful discussions. We are particularly
grateful to the anonymous referee for helpful comments.  This work is
supported by NASA through a Chandra Postdoctoral Fellowship award
PF3-40028 (ER-R).  GG-S acknowledged support from CONACYT 43121 grant.

\clearpage

\clearpage

\clearpage

\clearpage

\clearpage

\clearpage

\begin{thebibliography}{}

\bibitem[{Ballantyne & Ramirez-Ruiz (2001)}]{brr01} Ballantyne, D. R.,
\& Ramirez-Ruiz, E. 2001, ApJ, 559, L83

\bibitem[{Bloom et~al. (1999)}]{snbump} Bloom, J.~S., et~al.  1999,
Nature, 401, 453

\bibitem[{Bjorkman \& Cassinelli(1993)}]{bc93} Bjorkman, J. E.,
\& Cassinelli, J. P. 1993, ApJ, 409, 429

\bibitem[{Blandford \& McKee(1976)}]{bm76} Blandford, R.~D., \& McKee,
C.~F. 1976, Phys. Fluids, 19, 1130

\bibitem[{Blondin \& Lundquist(1993)}]{bl93} Blondin, J.~M., \&
Lundquist, P. 1993, ApJ, 405, 337

\bibitem[{Brown \& Bethe 1994}]{bb94} Brown, G.~E., \& Bethe,
H.~A. 1994, ApJ, 423, 659

\bibitem[{Castor et~al. (1975)}]{cmw75} Castor, J., McCray, R., \&
Weaver, R. 1975, ApJ, 200, L107

\bibitem[{Chevalier \& Liang(1989)}]{cl89} Chevalier, R. A., \& Liang,
E.~P. 1989, ApJ, 344, 332

\bibitem[{Chevalier \& Li(2000)}]{cl00} Chevalier, R. A., \& Li, Z.
2000, ApJ, 536, 195

\bibitem[{Chevalier et al. (2004)}]{cl00} Chevalier, R. A., Li, Z., \&
Fransson, C. 2004, ApJ, 606, 369

\bibitem[{Chiosi \& Maeder(1986)}]{cm86} Chiosi, C., \& Maeder, A.
1986, ARA\&A, 24, 329

\bibitem[{Clarke (1996)}]{cl96} Clarke, D. A. 1996, ApJ, 457, 291

\bibitem[{Contreras \& Rodriguez (1999)}]{cont99} Contreras, M. E.,
Rodriguez, L. F. 1999, ApJ, 515, 762

\bibitem[{Cowie \& McKee (1977)}]{cm77} Cowie, L.~L., \& McKee, C.~F.
1977, ApJ, 211, 135

\bibitem[{Dai \& Lu (2002)}]{dl02} Dai, Z.~G., \& Lu, T. 2002, ApJ, 565,
L87

\bibitem[{Dai \& Wu (2003)}]{dw03} Dai, Z.~G., \& Wu, X.~F. 2003, ApJ,
591, L21

\bibitem[{Djorgovski et~al. 2001)}]{djo02} Djorgovski, S.~G.,
et~al. 2001, ApJ, 562, 654


\bibitem[{Franco et~al.(1991)}]{f91} Franco, J., et~al. 1991, PASP,
103, 803

\bibitem[{Fransson et~al. (1996)}]{flc96} Fransson, C., Lundquist, P.,
Chevalier, R.~A. 1996, ApJ, 461, 993

\bibitem[{Fruchter et~al. (1999)}]{fruc} Fruchter, A. S., et al. 1999,
ApJ, 519, L13

\bibitem[{Galama et~al. (1998)}]{galama} Galama, T., et~al., 1998,
Nature, 387, 479

\bibitem[{Garc\'{\i}a-Segura \& Mac Low (1995a)}]{gsm95a} 
Garc\'{\i}a-Segura G., \& Mac Low M.-M. 1995a, ApJ, 455, 145

\bibitem[{Garc\'{\i}a-Segura \& Mac Low (1995b)}]{gsm95b}
Garc\'{\i}a-Segura G., \& Mac Low M.-M. 1995b, ApJ, 455, 160

\bibitem[{Garc\'{\i}a-Segura et~al.(1996a)}]{gs96} Garc\'{\i}a-Segura, G.,
Langer, N.,\& Mac Low, M.~M. 1996a, A\&A, 316, 133

\bibitem[{Garc\'{\i}a-Segura et~al. (1996b)}]{gs96b}
Garc\'{\i}a-Segura, G., Mac Low, M.~M., \& Langer, N. 1996b, A\&A,
305, 229

\bibitem[{Garc\'{\i}i-Segura \& Franco (1996)}]{gsf96} Garc\'{\i}a-Segura, G., 
\& Franco, J. 1996, ApJ, 469, 171

\bibitem[{Granot \& Loeb (2001)}]{gl01} Granot, J., \& Loeb, E. 2001,
ApJ, 551, L63

\bibitem[{Granot \& Ramirez-Ruiz (2004)}]{grr04} Granot, J., \&
Ramirez-Ruiz, E. 2004, ApJ, 609, L9

\bibitem[{Granot, Ramirez-Ruiz \& Loeb (2005)}]{grrl05} Granot, J.,
Ramirez-Ruiz, E., \& Loeb, A. 2005, ApJ, 618, 413

\bibitem[{Hartquist et~al. (1986)}]{hart} Hartquist, T.~W., et~al.
1986, MNRAS, 221, 715

\bibitem[{Heger et al. 2005}]{heger} Heger, A., Woosley, S.~E., \&
Spruit, H.~C. 2005, ApJ submitted (astro-ph/0409422)

\bibitem[{Heyl \& Perna 2003}]{hp03} Heyl, J.~S., \& Perna, R. 2003,
ApJ, 586, L13

\bibitem[{Hjorth et~al. (2003)}]{hjorth} Hjorth, J., et~al., 2004,
Nature, 423, 847

\bibitem[{Izzard, Ramirez-Ruiz, \& Tout(2004)}]{irt03} Izzard, R.~G.,
Ramirez-Ruiz, E., \& Tout C.~A., 2004, MNRAS, 348 1215

\bibitem[{Jakobsson et al. (2004)}]{jak04} Jakobsson, P., et al. 2004,
NewA,9, 435

\bibitem[{K\"onigl \& Granot(2002)}]{kg02} K\"onigl, A., \& Granot, J.
2002, ApJ, 574, 134

\bibitem[{Kwok et~al. (1978)}]{kpf78} Kwok, S., Purton, C. R., \&
Fitzgerald, P. M. 1978, ApJ, 219, L125

\bibitem[{Langer et al. (1988)}]{lan88} Langer N., Kiriakidis M., El 
Eid M.F., Fricke K.J., Weiss A., 1988, A\&A, 192, 177

\bibitem[{Langer(1991)}]{lan91} Langer, N. 1991, A\&A, 252, 669

\bibitem[{Langer(1998)}]{lan98} Langer, N. 1998, A\&A 329, 551

\bibitem[{Lipkin et al. (2004)}]{li04} Lipkin, Y. M., et al. 2004,
ApJ, 606, 381

\bibitem[{MacFadyen and Woosley(1999)}]{mw99} MacFadyen, A.~I., \&
Woosley, S.~E., 1999, ApJ, 524, 262

\bibitem[{MacFadyen et~al. (2001)}]{mwh01} MacFadyen, A.~I., Woosley,
S.~E., \& Heger, A. 2001, ApJ, 550, 410

\bibitem[{McKee (1986)}]{m86} McKee, C.~F.  1986, Ap\&SS, 118, 383

\bibitem[{M\'esz\'aros et~al. (1998)}]{mrw98} M\'esz\'aros, P., Rees,
M.~J., \& Wijers, R. 1998, ApJ, 499, 301

\bibitem[{Mirabal et~al. (2003)}]{mir03} Mirabal, N., et al. 2003,
ApJ, 595, 935

\bibitem[{Nakar et~al. (2003)}]{npg03} Nakar, E., Piran, T., \&
Granot, J. 2003, NewA, 8, 495

\bibitem[{Oren et al. (2004)}]{onp04} Oren, Y., Nakar, E., \& Piran,
T. 2004, MNRAS, 353, L35

\bibitem[{Ostriker and McKee (1988)}]{om88} Ostriker, J.~P., \& McKee,
C.~F. 1988, Rev. Mod. Phys., 60, 1

\bibitem[{Paczy\'nski(1998)}]{pac98} Paczy\'nski, B. 1998, ApJ, 494,
L45

\bibitem[{Panaitescu \& Kumar(2000)}]{pk00}Panaitescu, A., \& Kumar,
P.  2000, ApJ, 543, 66

\bibitem[{Panaitescu \& Kumar(2002)}]{pk01}Panaitescu, A., \& Kumar,
P.  2002, ApJ, 571, 779

\bibitem[{Piro \emph{et~al.}(2000)}]{pir00}Piro, L., et~al. 2000,
Science, 290, 955

\bibitem[{Podsiadlowski et al. (2005)}]{p05} Podsiadlowski, Ph.,
Mazzali, P. A., Nomoto, K., Lazzati, D., Cappellaro, E. 2004, ApJ,
607, L17

\bibitem[{Ramirez-Ruiz et~al. (2001)}]{rdmt01} Ramirez-Ruiz, E., Dray,
L., Madau, P., \& Tout, C.~A. 2001, MNRAS, 327, 829

\bibitem[{Ramirez-Ruiz, Merloni \& Rees (2001)}]{rmr01} Ramirez-Ruiz,
E., Merloni, A., \& Rees, M. J. 2001, MNRAS, 324, 1147

\bibitem[{Ramirez-Ruiz, Celotti \& Rees (2002)}]{rm02d} Ramirez-Ruiz,
E., Celotti, A., \& Rees, M.~J. 2002, MNRAS, 337, 1349

\bibitem[{Reeves et~al. (2002)}]{reev} Reeves, G. D.,  et~al. 2002,
Nature, 416, 512

\bibitem[{Rhoads (1997)}]{rh97} Rhoads, J. E. 1997, ApJ, 487, L1

\bibitem[{Salmonson (2003)}]{jay} Salmonson, J.~D. 2003, ApJ, 592,
1002

\bibitem[{Sari \& Piran(1995)}]{sp95} Sari, R., \& Piran, T. 1995,
ApJ, 455, L143

\bibitem[{Schaefer et al. (2003)}]{sch03} Schaefer, B. E., et
al. 2003, ApJ, 588, 387

\bibitem[{Soderberg \& Ramirez-Ruiz(2003)}]{sr03} Soderberg, A. M.,
Ramirez-Ruiz, E. 2003, MNRAS, 345, 854

\bibitem[Soberberg et~al. (2004)]{s04} Soderberg, A.~M., et~al.  2004,
Nature, 430, 648
 
\bibitem[{Stanek et~al. (2003)}]{snspec} Stanek, K. Z., et~al., 2003,
ApJ, 591, L17

\bibitem[{Stone \& Norman(1992)}]{sn92} Stone, J. M., \& Norman, M.~L.
  1992, ApJS, 80, 753

\bibitem[{Totani (2003)}]{tt03} Totani, T. 2003, ApJ, 598, 1151

\bibitem[{Trentham et~al.(2002)}]{trb02} Trentham, N., Ramirez-Ruiz,
E., \& Blain, A.~W. 2002, MNRAS, 334, 983

\bibitem[{Vietri \& Stella(1998)}]{vs98} Vietri, M., \& Stella,
L. 1998, ApJ, 507, L45

\bibitem[{Vishniac(1983)}]{vis83} Vishniac, E.~T. 1983, ApJ, 274, 152

\bibitem[{Wang et al. (2000)}]{wd00} Wang, X.~Y., Dai, Z.~G., \& Lu,
T. 2000, MNRAS 543, 90

\bibitem[{Wang \& Loeb (2000)}]{wl00} Wang, X., \& Loeb, A. 2000, ApJ
535, 788

\bibitem[{Waxman et~al. (1998)}]{waxman} Waxman, E., Frail, D., \&
Kulkarni, S. 1998, ApJ, 497, 288

\bibitem[{Weaver et~al.(1977)}]{w77} Weaver, R., et~al.  1977, ApJ,
218, 377

\bibitem[{Wijers et~al. (1998)}]{ralph} Wijers
R.~A.~M.~J., et~al. 1998, MNRAS, 294, L13

\bibitem[{Wijers(2001)}]{wij01} Wijers, R.~A.~M.~J. 2001, in Gamma Ray
Bursts in the Afterglow Era, ed. E. Costa, Frontera F. and Hjorth
J. (Springer: Berlin), 306

\bibitem[{Woltjer(1972)}]{wol72} Woltjer, L. 1972, ARA\&A, 10, 129

\bibitem[{Woosley(1993)}]{w93} Woosley, S.~E. 1993, ApJ, 405, 273

\bibitem[Zeh, Klose, \& Hartmann(2004)]{zkh04} Zeh, A., Klose, S., \&
Hartmann, D.~H.\ 2004, ApJ, 609, 952

\bibitem[{Zhang et~al. (2003)}]{zwm03} Zhang, W., Woosley, S.~E., \&
MacFadyen, A.~I. 2003, ApJ, 586, 356

\end{thebibliography}
\end{document}